\documentstyle[a4,epsfig]{article}
\newcommand{\be}{\begin{eqnarray}}
\newcommand{\ee}{\end{eqnarray}}
\newcommand{\nn}{\nonumber}
\newcommand{\ovl}{\overline}
\newcommand{\ra}{\rightarrow}

\newcommand{\ba}[1]{\begin{eqnarray} \label{(#1)}}
\newcommand{\ea}{\end{eqnarray}}

\newcommand{\dps}{\displaystyle}
\def\mNl{m_{\tilde{\nu}_\ell^N}}

\def\0n{0\nu\beta\beta}

\def\mN{m_{\tilde{\nu}_\ell^N}}
\def\mD{\tilde{m}_D}

\def \lsim {\mbox{${}^< \hspace*{-7pt} _\sim$}}

\def\missE{E \hspace{-0.65em}/}

\def\pmb#1{\setbox0=\hbox{#1}%
  \kern-.015em\copy0\kern-\wd0
  \kern.03em\copy0\kern-\wd0
  \kern-.015em\raise.0233em\box0 }

\def\pmb#1{\setbox0=\hbox{#1}
  \kern-.015em\copy0\kern-\wd0
  \kern.03em\copy0\kern-\wd0
  \kern-.015em\raise.0233em\box0}

\def\bm#1{\pmb{${#1}$}}

\begin{document}
\begin{center}
{\Large Sneutrino-induced like sign dilepton signal with conserved
R-parity}
\end{center}

\hspace{0.5cm}
\begin{center}
St. Kolb, M. Hirsch$^\dagger$, 
H.V. Klapdor-Kleingrothaus$^\ddagger$ 
and O. Panella 
\end{center}

\hspace{0.5cm}
\begin{center}
{\it Istituto Nazionale di Fisica Nucleare, 
Sezione di Perugia, Via A. Pascoli, I-06123 Perugia, Italy}\\
{\it $^{\dagger}$Department of Physics and Astronomy, 
University of Southampton, Highfield, Southampton SO17 1BJ, 
England}\\  
{\it $^{\ddagger}$Max-Planck-Institut f\"ur Kernphysik,
P.O. 10 39 80, D-69029 Heidelberg, Germany} 
\end{center}
\date{\today}

\hspace{0.5cm}
\begin{abstract}
Lepton number violation could be manifest in the sneutrino sector of 
supersymmetric extensions of the standard model with conserved R-parity. 
Then sneutrinos decay partly into the ``wrong sign charged lepton'' final 
state, if kinematically accessible. In sneutrino pair production or
associated single sneutrino production, the signal then is a like sign
dilepton final state.  Under favourable circumstances, such
a signal could be visible at the LHC or a next 
generation linear collider for a relative sneutrino mass-splitting 
of order ${\cal O}(0.001)$ and sneutrino width of order 
${\cal O}$(1 GeV). On the other hand, the like sign dilepton event rate 
at the TEVATRON is probably too small to be observable.

%
%

\end{abstract}


\section{Introduction}
Recently there has been special interest in the bosonic counterpart
of the neutrino appearing in supersymmetric (SUSY) extensions of the Standard
Model~\cite{susyphen}, the sneutrino. This is due to the intimate relation of 
neutrino and sneutrino properties as regards the violation of the lepton 
number L: either both neutrino and sneutrino violate L or both
do not~\cite{theorem,grosshab}. 
If L is violated, the sneutrino states in one generation exhibit 
a mass-splitting $\Delta m$=$m_1$-$m_2$, independent of the precise 
mechanism which generates L violation in the sneutrino sector.
The masses $m_{1,2}$ correspond to the mass states
$\tilde{\nu}^N_\ell$, $N$=1,2 (without mixing in generation space, 
$\ell$ denotes generation) which (in the effective low-energy theory)
are the physical states instead of the weak interaction states 
$\tilde{\nu}_L(\ell),\tilde{\nu}_L^*(\ell)$ (degenerate in mass in the
absence of L violation)~\cite{theorem,grosshab}. For the masses 
$m_{1,2}$ the relation $m_{1,2}$=$\overline{m}$$\pm$$\Delta m/2$ holds, 
where $\overline m$=$(m_1+m_2)/2$ is the average sneutrino mass.

Examples for the generation of a mass-splitting in the light sneutrino
sector include the SUSY sea-saw mechanism \cite{grosshab} and bilinear
R-parity violating models~\cite{rpviol}. 
In the sea-saw picture, below 
the electroweak symmetry breaking scale L violation is transferred from 
$SU(2)_L$ singlet sneutrinos to the light sneutrino sector. In models with 
bilinear R-parity breaking $SU(2)_L$ doublet sneutrinos mix with the Higgs 
sector and automatically violate L. 

The amount of L-violation (or equivalently $\Delta m$) is restricted by 
the kinematical
upper limits on neutrino masses and neutrinoless double beta decay 
~\cite{newdb,db}: the limit on $\Delta m$ is very stringent for
the first generation, but leaves room for appreciable L violation 
for the second and third generation (see the more detailed
discussion in section~\ref{lowenergycomp}). 
For example, the SUSY inverse of $\0n$ could be observable at a 
future muon collider~\cite{phen}. If 
the electroweak phase transition is weakly first 
or second order, a generation independent constraint on the mass 
splitting is set by baryogenesis~\cite{baryo}. 
%

A mass splitting 
also implies that the lightest sneutrino state may 
become the lightest supersymmetric particle (LSP) and - if its mass is 
around 70 GeV - a candidate for the cold dark matter in the 
universe~\cite{cdm}. This case should be easily discernable at 
a future lepton collider facility~\cite{cdmsig}. In the absence of L violation
the sneutrino cannot account for the cold dark matter in the 
universe~\cite{lightsneu,heavysneu}. 
%
%
In collider experiments, 
L violation in the sneutrino sector could manifest itself by 
a wrong sign lepton stemming from the decay of a sneutrino
in a lepton and a chargino~\cite{newdb,grosshab}: the electric charge of the 
lepton produced in the decay of sneutrinos (assumed heavier than charginos)
\be
\tilde{\nu}_{\ell}^N \ra \ell^{\pm} \chi^{\mp} 
\ee
may be of the wrong sign compared to the L-conserving case. In sneutrino
pair production or associated single sneutrino production mechanisms
the signal final state then contains a pair of like sign leptons 
(like sign dilepton or LSD). In the following, the LSD 
will be used as a signal for sneutrino-induced L violation and its 
cross section at proton and electron colliders will be estimated for 
different sneutrino production mechanisms under the assumption that 
R-parity is conserved. The LSD in a R-parity violating scenario 
(without sneutrino L violation) has been considered in 
ref.~\cite{rpvioldrei}. In view of the constraints on the mass-splitting 
mentioned above the discussion will be focussed on L-violation in the 
second and third generation.

The most obvious channel for sneutrino production is pair production:
$f \ovl{f} \ra \tilde{\nu}^1_\ell \tilde{\nu}_\ell^2$ in the $L$-violating 
case or $f \ovl{f} \ra \tilde{\nu}_L (\ell) \tilde{\nu}^*_L (\ell)$
in the L-conserving case. Here $f$ stands for either leptons or quarks.
At the TEVATRON (LHC) the pair production cross section for a 100 GeV 
sneutrino is around 30 fb (200 fb) \cite{sleptonlhc}, but drops quickly 
for larger values of the sneutrino mass: for a 300 GeV sneutrino the cross 
section is 0.03 fb (3 fb).    

On the other hand, at a future
electron positron Linear Collider (LC)~\cite{nlc} or at an accompanying 
electron-photon collider~\cite{ginzburg} the available energy in the 
center of mass system (c.m.s.) may
be substantially smaller than at the LHC and the sneutrino pair production 
may be not available. In this case single sneutrino production has to be 
considered in order to gain some insight into the sneutrino properties. 
Examples of processes, where a single sneutrino of any flavour is produced  
at the LC or an accompanying electron-photon collider are: 
\label{processes}
\begin{eqnarray} 
e^+ e^-   &\ra&  \tilde{\nu}_\ell^N \ell^{\pm} \chi^{\mp}_k 
\label{reactionone} \\
e^+ \gamma  &\ra&  \tilde{\nu}_\ell^N \ovl{\nu}_\ell 
\tilde{\ell}^+ \label{reactiontwo}\\ 
e^+ e^-   &\ra&  \tilde{\nu}_\ell^N W^{\pm} 
\tilde{\ell}^{\mp} \label{reactionthree} \ ,   
\end{eqnarray}
where $k$=1,2 and $\chi^{\pm}_2$ is the lighter of the two chargino states.
In R-parity violating frameworks, in addition to the mechanisms 
Eqs.~(\ref{processes}) single sneutrinos may be produced
in resonances or in two body final states both at hadron or lepton
colliders \cite{rpvioldrei,rpviolchem}.   

The processes in Eqs.~(\ref{reactionone},\ref{reactiontwo}) can be 
regarded as the leptonic SUSY analogues of single top-quark 
production~\cite{singletop} in the SM. It should also be recalled that 
sneutrinos can in 
addition be produced with accompanying neutrinos and neutralinos, {\it e.g.}
by the process $e^+ e^- \ra \tilde{\nu}_\ell^N \nu_{\ell} \chi^0_k$,  
resulting in a final state very difficult to observe or not observable
at all. Hence this possibility will not be considered further.

As the aim of this work is largely exploratory the attention is focussed 
on the value of total cross sections which are estimated 
at the energies the TEVATRON, LHC and a future $e^+e^-$ linear 
collider (LC). Kinematic cuts, which may be specific 
of the particular detector and/or experiment are not included. 
These can only be implemented in a realistic
MonteCarlo simulation of the signal with a parallel 
detailed study of the background which is however 
beyond the scope of this work. Nonetheless 
in order to have an idea of the observability of such phenomena 
the estimated total cross sections are compared to the (prospected) 
annual integrated luminosities of the TEVATRON, LHC and LC. The TEVATRON 
luminosity at $\sqrt{s}$=1.8 TeV is ${\cal L}_\circ$=2$fb^{-1}$,
the LHC luminosity at $\sqrt{s}$=14 TeV is expected to be
${\cal L}_\circ$=10$fb^{-1}$-100$fb^{-1}$, and the projected luminosity 
of the TESLA LC project is~\cite{wagner}  
$L_\circ$=300 fb$^{-1}$ at $\sqrt{s}$=500 GeV, 
and $L_\circ$=500 fb$^{-1}$ at $\sqrt{s}$=800 GeV. 
It is also expected that the LC will be able to operate
in the electron-photon ($e^-\gamma$), photon-photon
($\gamma\gamma$) and electron-electron ($e^-e^-$) 
modes with luminosities comparable to that of the 
underlying $e^+e^-$ machine 
(even higher in the case of the $\gamma\gamma$ mode)~\cite{telnov}. 

The outline of this note is as follows:
In the next section we start with an estimation of the LSD via
sneutrino pair production at the TEVATRON, the LHC and the LC. In 
section~\ref{sectionthree} the LSD produced in the reaction of  
Eq.~(\ref{reactionone}) is presented and in section~\ref{sectionfour} the 
LSD triggered by the process in Eq.~(\ref{reactiontwo})
will be considered. Since the reaction in Eq.~(\ref{reactionthree}) 
produces a more difficult signature due to the additional $W$ boson in the 
final state it will not be analysed. In 
section~\ref{lowenergycomp} the results are compared to low-energy 
constraints on $\Delta m$ and section~\ref{concl} contains the 
conclusions.


\section{The LSD in sneutrino pair production at TEVATRON, LHC and LC} 
\label{sectiontwo}

The graph contributing to the LSD generated by the Drell-Yan sneutrino 
pair production mechanism is depicted in Fig.~\ref{pairgraph}
and its amplitude is given explicitely in appendix A,
Eq.~(\ref{hadronamplitude}). The LSD is induced by the L-violating 
sneutrino propagator
~\cite{theorem}
\be \label{propagator}
\langle 0|T[\tilde{\nu}(x) \tilde{\nu}(y)]|0 \rangle =
\int \frac{d^4 k}{(2 \pi)^4} e^{-ik(x-y)}
\left[\frac{1}{m_1^2 - k^2 - i m_1 \Gamma_1} - 
\frac{1}{m_2^2 - k^2 - i m_2 \Gamma_2}\right] \ ,  
\ee
where $\Gamma_N$ is the total width of sneutrino mass state $N$.
In the small width approximation 
$1/(x^2 + \epsilon^2) \approx \pi \delta(x)/\epsilon$ (for
$\epsilon \ll x$) the LSD cross section becomes
\be \label{hadroncrosseq}
\sigma^{LSD}=\sigma(\tilde{\nu}_1 \tilde{\nu}_2) \ \xi^{pair} \ 
               \frac{\Gamma_1 (\chi^{\pm}l^{\mp})}{\Gamma_1}
               \frac{\Gamma_2 (\chi^{\pm}l^{\mp})}{\Gamma_2} \ ,
\ee
where $\Gamma_N(\chi^{\pm}l^{\mp})$ is the partial width of sneutrino
mass state $N$ into chargino final states. The effect of L violation in 
sneutrino pair production is measured by the factor
\be \label{xifactor}
\xi^{pair}  = 1-\frac{2m_1\Gamma_1 m_2\Gamma_2 
             (m_1\Gamma_1 m_2\Gamma_2+(\Delta m^2)^2)}
            {m_1^2\Gamma_1^2 m_2^2\Gamma_2^2+
             \left[(\Delta m^2)^2 + m_1^2\Gamma_1^2\right]
             [(\Delta m^2)^2 + m_2^2\Gamma_2^2]} \ , 
0 \leq \xi < 1 \ . 
\ee
The factor $\xi^{pair}$ is plotted in 
Fig.~\ref{pairxifig} for two sets of sample 
parameters defined in Table~\ref{tablesets} and for different values of 
$\ovl m$  as a function of the relative mass-splitting 
\be \label{defrelmass}
{\cal R}:=\frac{\Delta m}{\ovl m} \ : 
\ee
$\xi^{pair}$ is of order ${\cal O}$(1) for \({\cal R} \sim {\cal O}\)(0.01)
for moderate values of $\ovl m$, whereas, for large $\ovl m$, $\xi^{pair}$
is of order ${\cal O}$(1) only if ${\cal R}$ is of order ${\cal O}$(0.1). 

Integrating over the parton distribution functions, the cross section in  
Eq.~(\ref{hadroncrosseq}) to be expected at the TEVATRON and at the LHC 
is plotted in Fig.~\ref{lsdtevatron} and Fig.~\ref{lsdlhc}. 
(our results for the bare pair production cross section coincide with
the results of ref.~\cite{sleptonlhc}).

The total sneutrino width is estimated taking into account only two body decay 
modes
\be
\tilde{\nu}_{\ell}^{1,2} \ra \ell^{\pm} \chi_j^{\mp} \ , \ 
\tilde{\nu}_{\ell}^{1,2} \ra \nu \chi_j^0 .
\ee
since it is assumed that the sneutrino is heavier than the lighter
chargino. For the parameters chosen in Table~\ref{tablesets} the difference 
between the average sneutrino mass and the charged slepton mass is 
smaller than the mass of the weak gauge bosons. However, for a large 
mass-splitting the heavier sneutrino may decay as $\tilde{\nu}_{\ell}^1 \ra
\tilde{\ell}^{\pm}+W^{\mp}$. Since the charged slepton usually decays with
a large branching fraction into a neutralino-lepton final state, the
L-violating signal would be slightly enhanced. 

The final state charginos decay themselves mainly into the LSP 
neutralino and a fermion pair (to a small fraction charginos decay into 
four-fermion final states via gluinos or non-LSP neutralinos, see
the detailed discussion in ref.~\cite{charwidth}).
Therefore the signal final states are
\be \label{lhcsignature}
\mbox{LSD}^{\pm} + \ell^{\mp} + \ell^{`\mp}      + \missE \ , \
\mbox{LSD}^{\pm} + \ell^{\mp} + \mbox{ 2 jets }  + \missE \ , \ 
\mbox{LSD}^{\pm} + \mbox{ 4 jets }               + \missE \ . 
\ee
That is, if both charginos decay into leptons the final state may
be an exotic ``double'' LSD, {\it e.g.} $\tau^- \tau^- e^+ \mu^+$. 
In the SM, a LSD is produced in
the decays of $t \ovl t$, $b \ovl b$, $WW$, $WZ$ and $ZZ$ pairs or a 
singly produced $t$, {\it c.f.} the discussion in ref.~\cite{rpvioldrei}. 
The LSD in such events is, apart from a LSD in $ZZ$ production, always 
accompanied by jets, therefore the purely leptonic final states in 
Eq.~(\ref{lhcsignature}) remain unaffected by such a background. The
background from the decays of $ZZ$ pairs into four charged leptons of
same generation is vetoed by requiring $\missE \ > 2 m_{LSP}$.
%

As regards the semihadronic final states, provided that the velocity of the
decaying sneutrinos is large enough, 
the background event topology differs from the signal topology: hadron 
pairs decay into an odd number of jets on each side, while the number of 
jets in hadronic chargino decays is even; the LSD in the decays of a 
$WW$ pair is accompanied by four jets on one side, the LSD in a 
semihadronic decay of a $ZZ$ pair contains four jets on one and two 
jets on the other side, and the LSD in the decay of 
$WZ$ is accompanied by one single lepton. 

Conceivable SUSY background stems from decaying squark or gaugino pairs 
(in analogy to the SM background mentioned above) and decaying slepton
pairs. Here the LSD in decays of non-LSP neutralinos into two charged
lepton pairs may be eliminated rejecting events containing four
same-generation charged leptons. The LSD in the decay of a slepton pair 
is accompanied by a one-sided lepton pair in contrast to a single lepton
in the signal final state. In respect to semi-hadronic LSD production
channels, the same remarks as in the SM case apply. 
It therefore may be concluded that the signal is virtually background-free.


At the TEVATRON, even in favourable circumstances the LSD cross 
section (Eq.~\ref{hadroncrosseq})for the
parameter sets chosen is, at most, of order ${\cal O}$(0.1 fb), see 
Fig.~\ref{lsdtevatron}. Note that for large values of ${\cal R}$ and
small $\ovl m$ the LSD cross section drops to zero since the lighter
sneutrino cannot decay into the wrong sign charged lepton final
state. Hence the resulting event rate at TEVATRON is less than one 
event per year and detecting a sneutrino-induced LSD seems to be 
virtually impossible.

Given the larger \(\bm{pp}\) c.m.s. energy and luminosity the prospects of 
detecting a sneutrino-induced LSD certainly seem much brighter at the 
LHC. Since the LSD is virtually background-free, one LSD event per year
could be sufficient for detecting the effect. This in turn implies
that for a LHC-luminosity of ${\cal L}_\circ$=10$fb^{-1}$ 
(${\cal L}_\circ$=100$fb^{-1}$) a cross-section of 0.1fb (0.01fb)
yields an observable LSD signal. 
Therefore
for $\ovl m$ not much larger than 200 GeV the LSD is observable
(for both projected luminosities) for ${\cal R}$ being of order 
${\cal O}$(10$^{-3}$), see Fig.~\ref{lsdlhc}. Even for $\ovl m$ as 
large as 600 GeV the LSD could be visible if 
${\cal L}_\circ$=100$fb^{-1}$ and ${\cal R}$ being
of order ${\cal O}$(0.01) (without L violation the LHC may search for
sleptons not much larger than 350 GeV \cite{sleptonlhc}).
In section~\ref{lowenergycomp} the range
of ${\cal R}$ yielding an observable LSD signal is compared with 
the limits on $\Delta m$ from neutrino masses.

Finally, at a LC with a c.m.s. energy of 500 GeV (800 GeV) and for 
$\ovl m$ not much larger than 220 GeV (300 GeV), the LSD cross section 
is of order ${\cal O}$(0.01 fb) (yielding an annual event rate of order 
${\cal O}$(few/year)) for ${\cal R}$ of order ${\cal O}$(10$^{-3}$),
see Fig.~\ref{lsdnlc}. 

In section~\ref{lowenergycomp} the range
of ${\cal R}$ yielding an observable LSD signal at the LHC and LC
is compared with the limits on $\Delta m$ from neutrino masses.


\section{The LSD in the reaction 
$\bm{\lowercase{e}^+ \lowercase{e}^- \ra \chi^{\pm} 
\tilde{\nu}^N_\ell \ell^{\mp},  \ \ell=\mu,\tau}$}
\label{sectionthree}

The Feynman graphs contributing to this reaction 
are depicted in Fig.~\ref{feyntc} if the 
flavour of the final state sneutrino is either 
$\ell$=$\mu$ or $\ell$=$\tau$. If $\ell$=$e$, then further 
contributions similar to those examined in~\cite{elecsneu} 
(replacing the proton with an electron) have to be taken into account.
The resulting cross section can be written as
\be \label{eqtottc}
\sigma^{tot} = 
\sum_{k,l=\mbox{\footnotesize I};k \leq l}^{\mbox{\footnotesize III}} 
\sigma^{(k,l)} 
\ee 
where {\it e.g.} $\sigma^{\mbox{\footnotesize I,II}}$ is the contribution 
from the interference of $s$-channel gauge boson exchange 
(Fig.~\ref{feyntc}-I) and
$t$-channel sneutrino exchange (Fig.~\ref{feyntc}-II). 
The explicit expressions of the various contributions are listed
in appendix B. 


Since the graph in Fig.~\ref{feyntc}-III contains the derivative 
coupling of the $Z^0$ to the sneutrinos (see the previous section), 
its contribution is $\beta$-suppressed 
%
below the threshold of sneutrino pair production. For the sneutrino mass 
range considered in this section it is therefore 
%
always much smaller than the modes in
Figs.~\ref{feyntc}-I and~\ref{feyntc}-II and in the following 
it is neglected.

The three-particle phase space integral of graphs 
can be split into two two-particle phase space integrals 
(see {\it e.g.} ref.~\cite{BarPhi}), so that the phase space integrals
relative to the contributions in Fig.~\ref{feyntc}-I and~\ref{feyntc}-II 
(and their interference) can be evaluated analytically up to the 
integration over $Q^2$ to be integrated numerically, where $Q$ is the 
sum of the lepton and sneutrino impulses. We have checked, that using the
small width approximation 
the three body cross sections 
$\sigma^{\mbox{\footnotesize (III,III)}}$ and
$\sigma^{\mbox{\footnotesize (I,I)+(II,II)+(I,II)}}$ simplify in
factorised expressions of the sneutrino and chargino pair production times the 
corresponding sneutrino and chargino partial decay width 
if kinematically allowed.

If the final state sneutrino is lighter than at least one of the charginos,
the cross section depends very sensitively on the chargino width. However,
in what follows the sneutrino will taken to be heavier than both charginos
and their width will be neglected. In Fig.~\ref{mass2char} the cross
section Eq.~(\ref{eqtottc}) is plotted for the L-conserving case in 
terms of the final state sneutrino mass
for 
two sets of SUSY parameters $\mu,M_2,\tan \beta$ ($\mu$ is the Higgs
mixing parameter, $M_2$ is the gaugino mass associated to $SU(2)_L$, 
and $\tan\beta$ is the ratio of the vacuum expectation values of the 
two Higgs doublets) defined in 
Table~\ref{tablesets} and for beam energies $\sqrt{s}$=500 GeV and 
$\sqrt{s}$=800 GeV. In the following the 
unification condition in a minimal supergravity (mSUGRA) model 
$M_1$=(5/3)$M_2$$\times$tan$^2\Theta_W$ is assumed unless stated otherwise. 
In the sneutrio mass region where sneutrino pair production 
is kinematically excluded the cross section is of order ${\cal O}$(0.1 fb) 
or less. 

In order to estimate the cross section for the LSD, we use the
L-violating propagator Eq.~(\ref{propagator})
and again apply the small width approximation. The cross section 
Eq.~(\ref{eqtottc}) becomes:
\be \label{crosslviol}
\sigma^{LSD} & = & \xi^{single} \left[ \sigma^{tot}(\tilde{\nu}^1)
                   \frac{\Gamma_1(\chi^{\pm} \ell^{\mp})}{\Gamma_1}
                   \Theta(m_1^2-\bar{s}_-)\Theta(\bar{s}_+-m_1^2) 
\right. +  \nn \\ & & \left. \ \ \ \ \ \ \ \ \ \ 
                   \sigma^{tot}(\tilde{\nu}^2)
                   \frac{\Gamma_2(\chi^{\pm}\ell^{\mp})}{\Gamma_2}
               \Theta(m_2^2-\bar{s}_-)\Theta(\bar{s}_+-m_2^2) \right] \ ,
\ee
where 
\be
\bar{s}_+ = \sqrt{s}-m_{\chi_2}-m_{\ell} \ , \ \ 
\bar{s}_- = m_{\chi_1}+m_{\ell} . \nn 
\ee
The definition of $\bar{s}_-$ implies that the light sneutrino state 
contributes only if it is heavier than both charginos. Otherwise the
light sneutrino may be produced by the decays of a real chargino and
the width has to be taken into account (see the comment above). 
The L-violating factor $\xi^{single}$ measures the effect of L violation
in single sneutrino production and is defined as
\be \label{xidef}
\xi^{single} = 1-
               \frac{2 m_1 \Gamma_1 m_2 \Gamma_2}
                    {(\Delta m^2)^2 + m_1^2 \Gamma_1^2 + m_2^2 \Gamma_2^2}
\ , \ \ 0 \leq \xi < 1 \ .
\ee
The behaviour of $\xi^{single}$ in dependence on ${\cal R}$ is
illustrated in Fig.~\ref{xifigure}. For the parameters chosen
$\xi^{single}$ is of order ${\cal O}$(1) for ${\cal R}$ being
of order ${\cal O}$(0.1).  


The LSD signal final states in single sneutrino production are
the same as in pair production listed in Eq.~(\ref{lhcsignature}),
though the topology is different: the LSD lepton pair is one-sided,
whereas in pair production it is back to back. Conceivable SM and
SUSY background sources are the same as in sneutrino production, and
in prinicple the same remarks concerning their suppression apply. However,
the one-sidedness of the LSD in single sneutrino production is not 
inherent to any of the SM or SUSY background sources and itself 
provides a powerful criterium to distinct signal from background.
Therefore again it seems reasonable to conclude that the signal is
virtually background-free and we assume that an event rate of 
order ${\cal O}$(1/year) is sufficient to detect the effect so that
%
%
%
for an integrated luminosity of order 
${\cal O}$(few$\times$100 fb$^{-1}$) a cross section as small as 
${\cal O}$(0.01 fb) could be measured.

In Fig.~\ref{loglogfin} the cross section for the 
L-violating signal is plotted versus the relative sneutrino 
mass-splitting ${\cal R}$ 
for the parameter sets as defined in Table~\ref{tablesets} and
for different values of the average sneutrino mass $\ovl m$. 
The dominant contribution comes always from the lighter sneutrino state, 
and in extending the amount of L-violation it has been made sure that the 
lighter sneutrino is sufficiently 
heavy so that the widths of both charginos may be safely neglected. 
In the cases ($\ovl{m}$=275 GeV; $\sqrt{s}$=500 GeV; set A) and 
($\ovl{m}$=450 GeV; $\sqrt{s}$=800 GeV; sets A and B) a relative 
mass-splitting $\Delta m/\ovl{m}$ of order ${\cal O}$(0.01) is 
sufficient to produce an $L$-violating cross section of order 
${\cal O}$(0.01 fb). For the remaining cases 
($\ovl{m}$=350 GeV; $\sqrt{s}$=500 GeV; set A) and 
($\ovl{m}$=600 GeV; $\sqrt{s}$=800 GeV; sets A and B) the relative 
mass-splitting has to be one or two orders of magnitude larger in order
to produce a visible signal. The range of ${\cal R}$ yielding 
an observable LSD signal is compared to the low energy limits on 
$\Delta m$ in section~\ref{lowenergycomp}.



\section{The LSD in the reaction 
$\bm{\lowercase{e}^+ \gamma \ra \tilde{\nu}_\ell^N
          \tilde{\ell}^+\ovl{\nu}_{\lowercase{e}}, \ , (\ell=\mu,\tau)}$}
\label{sectionfour}

In electron-photon collisions $\mu$- and $\tau$-sneutrinos are
necessarily produced by graphs at least of order three in perturbation 
theory. The dominant contributions to higher generation
sneutrino production at an electron-photon collider are the ones 
depicted in Fig.~\ref{snunuslep}, which, as mentioned above,
are the leptonic SUSY equivalents
to single top-quark production in the SM at an electron-photon
facility~\cite{singletop} (the conventions adopted in this section
follow closely those of~\cite{singletop}). 
For the production of first generation slepton pairs additional 
contributions have to be taken into account, replacing for example
in Fig.~\ref{snunuslep}-III the $W$-boson by a chargino and
interchange the sneutrino and the neutrino, and similarly for 
the remaining contributions.
The relevant SUSY interaction Lagrangian are~\cite{susyphen}:
\be
{\cal L}_{W\tilde{\nu}\tilde{\ell}} & = & 
   \frac{-i g}{\sqrt{2}} W^+_{\mu}(\tilde{\nu}_{\ell,L}^* 
   \stackrel{\leftrightarrow}{\partial}^{\mu}\tilde{\ell}_L)+h.c. \ , 
\nn \\
{\cal L}_{\gamma\tilde{\ell} \tilde{\ell} } & = & 
     i e A_{\mu} \ \tilde{\ell}_{L,R}^*
          \stackrel{\leftrightarrow}{\partial}^{\mu}\tilde{\ell}_{L,R} \ ,
\nn \\
{\cal L}_{\gamma W \tilde{\ell} \tilde{\nu}} & = &
     \frac{g}{\sqrt{2}} e A^{\mu} \tilde{\ell}_L^* \tilde{\nu}_{\ell,L} 
W_{\mu}^+ 
     + h.c. \ .
\ee  
The gauge invariant amplitude consists of four terms and can be written
as
\be
{\cal M}=
\frac{i e g^2}{2}
\sum_{i=\mbox{\footnotesize I}}^{\mbox{\footnotesize IV}} 
T_{\mu}^{i}\epsilon^{\mu}(p_g) \ . 
\ee
The various contributions are listed in Appendix C.
Here it is not possible to factorize the phase space integral into two
two-particle integrations as it was done in the previous section
so that the phase space integration is performed numerically using the
VEGAS~\cite{lepage} code. Following~\cite{Byk}, in the frame 
$\vec{\bf p}_{\tilde{\nu}}=-\vec{\bf p}_{\tilde{\ell}}$ the total 
cross section is given  by  
\be \label{egammacross} 
\sigma_{tot}(s)=\frac{1}{1024 \pi^4}
\int\limits_{(m_{\tilde{\nu}_\ell^N}+m_{\tilde{\ell}})^2}^s d s_2 
\int\limits_{-(s-s_2)}^0 d t_1 
\int d\cos\theta\, d \phi 
\frac{\lambda^{1/2}(s_2,m_{\tilde{\nu}_\ell^N}^2,m_{\tilde{\ell}}^2)}{s_2}
|{\cal M}|^2 \ , \nn \\
\ee
where $s_2=(p_{\tilde \ell}+p_{\tilde{\nu}_\ell^N})^2$ and 
$t_1=(p_e-p_{\nu})^2$. The explicit parameterization of the individual 
four-vectors in terms of the invariants $s_2,t_1$ and the polar angles of 
$\vec{\bf p}_{\tilde{\ell}}$ is given in appendix D. 

However the $e\gamma$ option of a LC will be realized producing high 
energy photons through Compton-backscattering of 
a low energy laser beam with  an high energy positron 
beam~\cite{ginzburg}. Thus the resulting photons are not 
monochromatic and one has to fold the cross section of any  $e\gamma$ process 
over a photon energy spectrum. It is expected that a photon collider will 
operate at luminosities very close to that of the $e^+e^-$ machine. 
In terms of the variables 
\be 
x  =  \frac{4 E_{e^+} E_{Laser}}{m_e^2} \leq 2(1+\sqrt{2})\ \ , \ \ 
y  =  \frac{E_{\gamma}}{E_{e^+}} \nn
\ee
the photon energy spectrum is given by~\cite{Gho98}
\be \label{folding}
{\cal P}(y)  =  \frac{1}{N}
   \Big[1-y+\frac{1}{1-y}-\frac{4y}{x(1-y)}+
        \frac{4y^2}{x^2(1-y)^2}\Big] \ . 
\ee
Here the factor
\be
N  =  \frac{1}{2}+\frac{8}{1+x}+\frac{7}{2x(1+x)}+
        \frac{1}{2x(1+x)^2}+(1-\frac{4}{x}-\frac{8}{x^2})\ln(1+x)\nn
\ee
normalizes $\int {\cal P}(y)dy$ to unity. The resulting
cross section applying Eq.~(\ref{folding}) is
\be \label{foldedcross}
\sigma=\int\limits_{(m_{\tilde{\nu}^N_\ell}+m_{\tilde{\ell}})^2/s}^{x(x+1)}
 \  {\cal P} (y) \  \sigma_{tot} (y s) \ dy \ .
\ee
where $\sigma_{tot}$ is defined in Eq.~(\ref{egammacross}).
Assuming the mSUGRA relations for the slepton masses~\cite{SUGRAmass}
\be
m_{\tilde{\nu}_L}^2 & = &
M_0^2 + 0.07 m_{\tilde{g}}^2 + \frac{1}{2}\cos 2 \beta M_Z^2 \nn \\
m_{\tilde{\ell}_L}^2 & = &
M_0^2 + 0.07 m_{\tilde{g}}^2 + \frac{1}{2}\cos 2 \beta M_Z^2
(2 \sin^2 \Theta_W -1) \nn 
\ee 
The cross section Eq.~(\ref{egammacross}) is plotted in
Fig.~\ref{bare} for the parameters defined in Table 
\ref{tablesets} and for two different  values of the center-of-mass energy of
the $e^+ e^-$-pair ($x$ was set to its maximum). The cross section is
of order ${\cal O}$(0.1 fb) at $\sqrt{s}$=800 GeV and $M_0$ masses of 
order ${\cal O}$(100 GeV). For a more realistic center of mass energy 
of $500$ GeV the resulting cross sections are about an order of
magnitude smaller. With luminosities as those projected for a LC of 
several $10^{34}$ cm$^{-2}$s$^{-1}$ one produces several tens 
of single sneutrino events per year.

The process 
\(\lowercase{e}^+ \gamma \ra \tilde{\nu}_\ell^N \tilde{\ell}^+
\ovl{\nu}_{\lowercase{e}}, \, (\ell=\mu,\tau)\) can also be exploited at
$e^+ e^-$ colliders. Here infact by taking the photon to be virtual 
the diagrams of Fig.~\ref{snunuslep} 
can be attached to the electron current thus describing 
slepton associated single sneutrino production at $e^+e^-$ colliders 
The full process involving virtual photons (with possibly other diagrams)
due to the fact that the photon is massless (its propagator gives a factor 
$Q^{-4}$ in the squared amplitude if $Q$ is the virtual photon momentum) 
is dominated by small values of $Q$ i.e. 
quasi real photons and the full $e^+e^-$ process is approximated by 
folding the cross section of Eq.~(\ref{egammacross}) with a photon 
distribution function $f_\gamma(x)$. This is the 
so-called equivalent photon (Weisz\"acker-Williams) approximation (E.P.A) 
which, quite generally, for any scattering process involving a charged 
particle in the initial state that interacts via a virtual photon, 
consist in approximating the full process by defining  a distribution
function for the photon in the electron of energy $E$.
The Weisz\"acker-Williams spectrum is given by~\cite{ter} 
\be
f_{\gamma}(x) = \frac{\alpha}{\pi} \log\left(\frac{E}{m_e}\right)
                \frac{1+(1-x)^2}{x}  \nn
\ee
and $f_{\gamma}(x)$ is interpreted as the photon distribution of 
an electron of energy $E$ i.e. the probability 
that an electron of energy $E$  radiates a quasi-real 
photon in the forward direction with energy $E_\gamma=xE$. 
The cross section at the $e^+e^-$ machine is estimated using the following 
formula:
\be
\sigma  =   \int\limits_{(m_{\tilde{\nu}^N_\ell} + m_{\tilde{\ell}})^2/s}^1 
           f_{\gamma}(x)\ \sigma_{tot}(x s) \ dx \ . 
\label{photonapprox} 
\ee
Some examples of the results employing Eq.~(\ref{photonapprox}) for 
the production of second and third generation
single sneutrinos are shown in Fig.~\ref{bare}. Again, for plausible
values of slepton masses and center-of-mass energies the resulting
total cross sections are of the order of
${\cal O}(0.1 fb)$. Therefore, as in the production
of first generation sneutrinos (see ref.~\cite{sneuchar})
the production of second and third generation 
sneutrinos at an $e^+e^-$ collider through the $e\gamma$ subprocess
appears to be at observable rates provided that the  LC will operate
at the prospected luminosities of several $10^{34}$ cm$^{-2}$ s$^{-1}$.

Applying the procedure of the previous section, it is possible to
estimate the rate of wrong charged lepton events coming from
L-violating sneutrino decays. 
If the initial state lepton is a positron, the signal final
states are (assuming the final state leptons being not too soft) 
\be \label{egammasignallsd}
\mbox{LSD}^+ + \chi^- + \missE \ \
\mbox{(for $\tilde{\ell}^+ \ra \ell+ \chi^0$)} \ ,  
\ee
\be \label{egammasignalwscl}
\ell^+ + \chi^- + \chi^+ + \missE \ \ 
\mbox{(for $\tilde{\ell}^+\ra\chi^+\ovl{\nu}_{\ell}$)} 
\ ,
\ee  
and the possible chargino decay modes have been mentioned
in section~\ref{sectiontwo}.
In the SM the second or third generation LSD in Eq.~(\ref{egammasignallsd})
may be 
produced in $e$-$\gamma$-collisions in at least 7th order of 
perturbation theory in the decay chain of a top-bottom final state 
(with additional CKM suppression). The rate of such a process is negligible 
even compared to the very low signal rate. The 
wrong charged sign lepton 
in Eq.~(\ref{egammasignalwscl}) may be mimicked by the production of 
a $Z^0$-$e^+$ pair (the $Z^0$ decaying into $\ell$'s), when the
charginos to a large fraction decay into the $\ell^+ \nu_{\ell} \chi^0$
final state. The same reasoning holds for the SUSY background: 
{\it e.g.} the final state containing two charginos may be reproduced
by the decays of a selectron and a non-LSP neutralino
\be 
e^+ \gamma \ra \tilde{e}^+ \chi^0 \ , 
\tilde{e}^+ \ra \chi^+ \ovl{\nu}_e \ ,
\chi^0 \ra \chi^- \ell^+ \nu_{\ell} \ . \nn
\ee
If the $\tilde{\ell}$ decays to a large fraction directly into the LSP
(thus producing a $\ell^+ \ell^+$ final state) 
the background may be eliminated discarding events with 
only one $\ell^+$ in the final state. In the opposite case, a detailed 
analysis of the angular and energy distribution of the final state 
particles is required.

The resulting cross section for the LSD signal and the wrong sign
charged lepton signal is plotted in Fig.~\ref{egamma} for
different combinations of parameters. 
In the examples chosen, for a moderate ratio 
\({\cal R} \sim 10^{-2}\) the L-violating signal 
may be observable only if both sleptons and charginos are light  
($M_0 \approx M_2 \approx |\mu| \approx 100$ GeV ).
For higher masses an appreciable signal is reached only for a large
amount of L-violation.

Note that in mSUGRA the $SU(2)_L$-doublet charged sleptons are 
slightly heavier than the $SU(2)_L$-doublet sneutrinos. This fact
renders the process discussed here less favourable than the process
discussed in the previous section. Larger cross sections for  
single $\mu$- and $\tau$-sneutrino production may be possible in 
scenarios where charged sleptons are allowed to be (considerably) 
lighter than the sneutrinos. Two examples shown in 
Fig.~\ref{noSUGRA} illustrate this fact. For the parameters chosen,
L-violating cross sections of order ${\cal O}$(0.01 fb) can be 
obtained for \({\cal R} \sim 0.01\) for higher values
of $\ovl{m}$ in respect to the mSUGRA case in Fig.~\ref{egamma}.
In the next section the range of ${\cal R}$ rendering an observable
L-violating signal is compared to the low-energy constraints on 
$\Delta m$.

\section{Observability of the LSD SIGNAL in view of low--en\-er\-gy 
con\-straints on \(\bm{\Delta \lowercase{m}}\)}
\label{lowenergycomp}

The results for the LSD signal obtained in the previous sections
should be compared to the
low-energy limits on the sneutrino mass-splitting. The most stringent
bounds on $\Delta m$ come from the smallness of neutrino masses: loops 
containing L-violating sneutrinos and neutralinos contribute to the Majorana 
neutrino mass matrix ${\cal M}_{LL}$ mixing the $SU(2)_L$ doublet fields 
~\cite{newdb,db,grosshab}. In ref.~\cite{db} a
scan over a large range of the SUSY parameter space has been carried
out and the following ``absolute'' limit has been derived from the
upper limits on neutrino masses:
\be \label{neutrlimits}
\Delta m (\ell) < 156 \, 
\frac{m_{\nu}^{exp}(\ell)}{1 \hbox{ eV}} \hbox{ keV} \ ,
\ee
where $\ovl{m}$ has been set to 100 GeV. Higher values for $\ovl{m}$
result in less stringent bounds.
This in turn implies that: \((i)\) a relative mass-splitting 
\({\cal R} \sim 10^{-2}\) (yielding observable LSD event rates in sneutrino
pair production at the LHC and a LC and in single sneutrino production
at a LC) is compatible with a neutrino mass of order
\({\cal O}(10 \hbox{keV}\)) and \((ii)\) \({\cal R}\) 
being of order \({\cal O}(10^{-3})\)
(yielding observable LSD event rates in sneutrino pair production at
the LHC and a LC) is compatible with a neutrino mass of order 
${\cal O}$(1 keV). It is interesting to note that a neutrino with
a mass of order ${\cal O}$(1 keV) could provide warm dark matter
without overclosing the universe~\cite{warmdm}. 
%
%
The current kinematical 
limits on neutrino masses are~\cite{pdg}
\be
m_{\nu}^{exp}(e) < 3 \hbox{ eV} 
\ , \ 
m_{\nu}^{exp}(\mu) < 190 \hbox{ keV} 
\ , \ 
m_{\nu}^{exp}(\tau) < 18.2 \hbox{ MeV} \ , \nn
\ee 
{\it i.e.} the kinematical limits leave room for an observable sneutrino
mass-splitting in LSD events in the second and third generation.
 
On the other hand, 
the limit on the $m_{ee}$ entry of ${\cal M}_{LL}$ from the 
Heidelberg-Moscow $\0n$-decay experiment~\cite{baudis} together with the 
preliminary results of neutrino oscillation experiments (for a
recent overview see {\it e.g.}~\cite{oscill}) imply that all entries 
of ${\cal M}_{LL}$ should satisfy $m_{ij}\, \lsim \, 2.5 $ eV~\cite{paes}.
If the oscillation solution (requiring mass-squared differences between 
the neutrino states of order ${\cal O}$(1 eV$^2$) or less) will be
confirmed both 
for the atmospheric and for the solar neutrino problem, sneutrino-induced 
L violation will not be directly visible for sneutrino decay widths of 
order ${\cal O}$(1 GeV). If, on the other hand, the oscillation solution to
one or both of the neutrino problems 
are not confirmed, several entries of ${\cal M}_{LL}$ are no longer 
required to be small, and neutrinos could still be much heavier than
currently accepted. 
 
Furthermore, it has been pointed out in ref.~\cite{grosshab} 
that sneutrino oscillations could be observable even 
for neutrino masses of order ${\cal O}$(1 eV) provided that the total 
decay widths of the sneutrino mass states are very small. In the approach
taken here this means that $\xi$ is of order ${\cal O}$(1) 
for very small values of $\Delta m$. As an example, this fact is
illustrated for associated sneutrino production at a future LC: 
for a neutrino mass 
of 1 eV (corresponding to $\Delta m$=156 keV) the L-violating cross
section for set A and $\sqrt{s}$=500 GeV is
of order ${\cal O}$(0.01 fb) provided that the sneutrino widths are of 
order ${\cal O}$(1 MeV), see Fig.~\ref{varywidth} where the sneutrino width 
has been varied freely. Such tiny sneutrino 
widths are conceivable for a small mass difference of sneutrinos in respect 
to the lighter chargino and the LSP neutralino (still it has to made sure
that the final state leptons are not too soft to be detectable). This is 
usually impossible in a mSUGRA context, where the LSP is considerably 
lighter than the lighter chargino, but in a more general context may well
be possible. If {\it e.g.} $M_1=M_2$, it is always possible 
to find regions in the SUSY parameter space where the sneutrino width 
becomes very small and L violation may be directly visible.

\section{Conclusions}\label{concl}


In conclusion, L violation in the sneutrino sector of supersymmetric
extensions of the Standard Model may manifest itself by decays of 
sneutrinos into final states containing a like sign dilepton. Since the LSD 
signal is virtually background-free even very small event rates of order 
${\cal O}$(1/year) could be observable at hadron, electron or 
electron-photon collider facilities.  

At the LHC or a future electron linear collider (LC) with a c.m.s. 
energy of 800 GeV, 
an observable LSD is triggered in sneutrino pair production for a relative 
mass-splitting ${\cal R}$ being of order ${\cal O}$(10$^{-3}$) if the average 
mass of the two sneutrino mass states $\ovl m$ is not much larger than 400 GeV.
At a LC with a c.m.s. energy of 500 GeV the LSD is observable for 
\({\cal R}\) being of order \({\cal O}(10^{-3}\)) 
if \(\ovl{m}\ \lsim\ 220\) GeV. 
At the TEVATRON, the LSD event rate is less than one per year for realistic
SUSY parameters and therefore virtually not observable. 

If $\ovl m$ is larger than the c.m.s. energy available ({\it e.g.} at 
500 GeV next linear collider), sneutrino pair production is excluded and
associated single sneutrino production has to be considered in 
future collider facilities such as a LC or an electron-photon machine. 
For chargino-associated single sneutrino prodution at a
LC, under favourable cirumstances (large branching fraction of the
chargino-lepton decay channel and a relative low average sneutrino
mass) one may expect about ten LSD events per year 
for ${\cal R}$ being of order ${\cal O}$(0.01). 
Under unfavourable circumstances less than one 
LSD event per year is expected. Then an appreciable event rate is possible 
only for ${\cal R}$ being of order ${\cal O}$(1). 

For charged slepton associated sneutrino production at an 
electron-photon collider, in mSUGRA models the event rate of the LSD or
the wrong charged lepton signal is very low for realistic slepton masses 
and ${\cal R}$ being of order ${\cal O}$(0.01). In order to produce a
detectable event rate ${\cal R}$ must be close to
one. If the charged slepton is considerably lighter than the average
sneutrino mass, an observable event rate is possible for moderate
values of the mass-splitting.

L violation in the sneutrino sector induces radiative contributions
to the left-handed Majorana mass-matrix ${\cal M}_{LL}$: 
${\cal R}$ being of order 
${\cal O}$(10$^{-3}$) corresponds to a neutrino mass of order
${\cal O}$(1 keV), ${\cal R}$ being of order
${\cal O}$(10$^{-2}$) corresponds to a neutrino mass 
of order ${\cal O}$(10 keV). Such an amount of L violation 
is compatible 
with the kinematical limits on second and third generation neutrino, 
but not compatible with the limits on the entries of 
${\cal M}_{LL}$ from the neutrinoless double beta decay combined
with the oscillation solution to the solar and atmospheric neutrino
problems.\\

\begin{center}
{\large \bf Acknowledgments}
\end{center}
O.~P. would like to thank the Max-Plank-Institut f\"ur Kernphysik, 
for the very warm hospitality during his stay in Heidelberg. The work
of St.~K. was in part supported by a Marie Curie Individual Fellowship
(European Union), under contract N. HPMF-CT-2000-00752.
 
\appendix
\section{LSD in sneutrino pair production}
\label{hadronappendix}
The amplitude for the LSD in sneutrino pair production at a hadron
collider is (see Fig.~\ref{pairgraph})
\be \label{hadronamplitude}
{\cal T}_{fi} & = & \frac{g^4}{4 \cos^3 \Theta_W} V_{i1}V_{i'1}
|{\cal D}_Z (s)|^2 (p_{34}-p_{12})
\cdot j^{quark} j^{lep}_{il} j^{lep}_{i'l'} \nn \\
              & \times & 
({\cal D}_1 (s_{34}){\cal D}_2 (s_{12})-
 {\cal D}_1 (s_{12}){\cal D}_2 (s_{34})) \ .
\ee
Here $j$ are the quark and lepton currents and propagator factors are
denoted as ${\cal D}(x)=(m^2 - x - i \Gamma m)^{-1}$. The invariants
are $s_{12} = p_{12}^2 = (p_i + p_l)^2$, 
    $s_{34} = p_{34}^2 = (p_i' + p_l')^2$ an $V$ is the
chargino mixing matrix. Note that Eq.~(\ref{hadronamplitude}) 
reflects the off-diagonality of the sneutrino-$Z$ vertex.

\section{Contributions to 
\bm{\lowercase{e}^+ \lowercase{e}^- \ra \chi^{\pm} 
\tilde{\nu}^N_\ell \ell^{\mp}, \, \ell=\mu,\tau}}

In the following individual contributions to the cross section for the 
reaction  $e^+ e^- \ra \chi^{\pm} \tilde{\nu}^N_\ell \ell^{\mp}$ 
(Fig.~\ref{feyntc}) are listed for $\ell=\mu,\tau$. 
For $\ell=e$ further contributions have to be taken into account. 
The contribution of the Higgs exchange diagram has been neglected
since it is proportional to the electron mass. The index $i$ corresponds
to the final state chargino (for obtaining numerical results only 
production of the lighter chargino was considered), the indices $j,j'$ 
correspond to exchanged charginos. $V,V'$ stands for either $\gamma$ or 
$Z^0$ and $N$
is the index for the heavier ($N$=1) or lighter ($N$=2) sneutrino
mass states. $m_D$ is the usual $L$-conserving sneutrino 
mass. 
The modification of the sneutrino propagator in the presence of $L$ violation
has been neglected. Propagator factors 
$1/\left[(M^2-X)^2+M^2 \Gamma_M^2\right]$
are denoted by $P(M,X)$. The integration limits of 
$Q^2=(p_{\mN} + p_\ell)^2$ are 
\be
Q^2_{max,min}=(\sqrt{s}-m_i)^2, (m_\ell+\mNl)^2\nn
\ee
unless noted otherwise. The total cross section is a sum of the
contributions depicted in Fig.~\ref{feyntc} and their 
interferences. The interference between~\ref{feyntc}-III and the
other graphs is neglected since in the numerical examples displayed
in Table~\ref{tablesets} 
the contribution of~\ref{feyntc}-III
is much smaller than the contribution from both~\ref{feyntc}-I 
and~\ref{feyntc}-II.

The contribution from ($t$-channel) sneutrino-exchange shown in
Fig.~\ref{feyntc}-I is defined as 
\be
\sigma^{(\mbox{\footnotesize I,I})}_i & = &
       \frac{1}{2s}\frac{1}{4}\sum_{j,j'}g^6|V_i|^2|V_j|^2|V_{j'}|^2
       \frac{1}{128 \pi^3} 
       \int d Q^2 \;  
       \lambda^{1/2}(1,\frac{m_\ell^2}{Q^2},\frac{\mNl}{Q^2})
       P(m_j,Q^2)
\nn \\ \nn \\
&\times& P(m_{j'},Q^2) (Q^2+m_\ell^2-\mNl^2)
   ((m_j^2-Q^2)(m_{j'}^2-Q^2)+m_j m_{j'}\Gamma_j \Gamma_{j'})
\nn \\ \nn \\
&\times& 
   \Big[ \lambda^{1/2}(\frac{Q^2}{s},\frac{m_i^2}{s},1)
   (1+\frac{(Q^2-\mN^2)(m_i^2-\mN^2)}{(Q^2-\mN^2)(m_i^2-\mN^2)+\mN^2 s})
\nn \\ \nn \\
& + & \frac{1}{s}(2 \mN^2-m_i^2-Q^2)\ln F(Q^2) \Big] \ , \nn
\ee
where as usual $\lambda(a,b,c)$=$a^2+b^2+c^2+2ab+2ac+2bc$ and
the function $F(Q^2)$ is defined as
\be \label{fqsq}
F(Q^2)& = & (m_i^2-\mD^2+\frac{1}{2}(Q^2-m_i^2-s)+
       \frac{1}{2}\lambda^{1/2}(Q^2,m_i^2,s)) \nn \\ \nn \\
& \times &         (m_i^2-\mD^2+\frac{1}{2}(Q^2-m_i^2-s)-
       \frac{1}{2}\lambda^{1/2}(Q^2,m_i^2,s))^{-1} \ . \nn
\ee

The contribution from ($s$-channel) gauge-boson-exchange depicted
in Fig.~\ref{feyntc}-II is defined as
\be
\sigma^{(\mbox{\footnotesize II,II})}_i & = & 
     \frac{1}{2s}\frac{1}{4} \sum_{j,j'}\sum_{V,V'}
       g^2 V_{j1}V_{j'1} (g^L_V g^L_{V'} + g_V^R g_{V'}^R)
      \frac{1}{32 \pi^3} \int d Q^2\;
      \lambda^{1/2}(1,\frac{m_\ell^2}{Q^2},\frac{\mNl^2}{Q^2})
    \nn \\ 
&\times& (Q^2+m_\ell^2-\mN^2) \lambda^{1/2}(1,\frac{m_l^2}{s},\frac{Q^2}{s})
        P(V,s)P(V',s)P(j,Q^2)P(j',Q^2) 
    \nn \\ \nn \\
&\times& G(Q^2) \Big\{ \frac{1}{3} s^2 \Big[ 
      \lambda(1,\frac{m_i^2}{s},\frac{Q^2}{s})
   + (1+\frac{m_i^2}{s}+\frac{Q^2}{s}-\frac{2(m_i^2-Q^2)^2}{s^2})\Big]
    \nn \\ \nn \\
&\times& \Big[ Q^2 (O^{'L}_{ji})_V (O^{'L}_{j'i})^*
       +m_j m_{j'} (O^{'R}_{ji})_V (O^{'R}_{j'i})^*_{V'} \Big]
\nn \\ \nn \\ &+&2Q^2 s \Big[m_{j'} m_i (O^{'L}_{ji})_V (O^{'R}_{j'i})^*_{V'}
            + m_{j'} m_i (O^{'R}_{ji})_V (O^{'L}_{j'i})^*_{V'}
 \Big] \Big\} \ , \nn 
\ee
where the function $G(Q^2)$ is given by
\be
G(Q^2)&=&\Big[ (m_V^2-s)\{(m_{V'}^2-s)a-m_{V'}\Gamma_{V'}b\} \nn \\
      &+& m_V \Gamma_V\{m_{V'}\Gamma_{V'}a+(m_{V'}^2-s)b\} \Big]\nn \ ;\\
a&=&(m_j^2-Q^2)(m_{j'}^2-Q^2)+m_j m_{j'} \Gamma_j \Gamma_{j'}\nn \\
b&=&m_{j'}\Gamma_{j'}(m_j^2-Q^2)-m_j\Gamma_j (m_{j'}^2-Q^2) \ . \nn
\ee
The lepton-gauge boson couplings $g^{L,R}_V$ are
\be
g^L_{Z^0} & = &\frac{g}{\cos\Theta_W}(\frac{1}{2}-\sin^2\Theta_W) \ \ , \ \ 
g^R_{Z^0}=\frac{g}{\cos\Theta_W}(-\sin^2\Theta_W) \nn \\ \nn \\
g_{\gamma}^L & = & g_{\gamma}^R=e \ \ , \nn
\ee
and the gauge boson-chargino couplings $(O^{'L,R}_{ij})_V$ are
\be
(O^{'L}_{ij})_{Z^0}&=&\frac{g}{\cos\Theta_W}
    (-V_{i1}V_{j1}^*-\frac{1}{2}V_{i2}V_{j2}^*+\delta_{ij}\sin^2\Theta_W)
\nn \\ \nn \\ 
(O^{'R}_{ij})_{Z^0}&=&\frac{g}{\cos\Theta_W}
    (-U_{i1}^*U_{j1}-\frac{1}{2}U_{i2}^*U_{j2}+\delta_{ij}\sin^2\Theta_W)
\nn \\ \nn \\
(O^{'L}_{ij})_{\gamma}&=&(O^{'R}_{ij})_{\gamma}\;=\;-e\delta_{ij} 
\nn
\ee
(Note that $(O_{ij}^{'L,R})_V=(O_{ji}^{'L,R})_V^*)$.

The interference between sneutrino and gauge boson graphs is
\be
\sigma_i^{(\mbox{\footnotesize I,II})} & = &
      \frac{1}{2s}\frac{1}{4}\sum_{j,j'}g^4
       |V_{i1}|^2|V_{j1}|^2\frac{1}{32\pi^3}\int d Q^2\;
       (Q^2+m_l^2-\mN^2)
    \lambda^{1/2}\Big( 1,\frac{m_\ell^2}{s},\frac{Q^2}{s} \Big)
    \nn \\ 
&\times&  P(j,Q^2)P(j',Q^2)P(V,s)
   \Big\{(m_V^2-s)\Big[(m_j^2-Q^2)(m_{j'}^2-Q^2)+m_j m_{j'}
   \Gamma_j\Gamma_{j'} \Big]
   \nn \\ 
&-& m_V \Gamma_V \Big[ (m_j^2-Q^2)m_{j'}\Gamma_{j'}-
        (m_{j'}^2-Q^2)m_j\Gamma_j \Big]
\Big\}
   \nn \\ \nn \\
& \times &
 \Big\{m_i m_{j'}(O^{'R}_{i j'})^*_V(-\ln F(Q^2))
       +\frac{1}{2}(O^{'L}_{ij'})_V^* \Big[\lambda^{1/2}
      \Big( \frac{Q^2}{s},\frac{m_i^2}{s},1 \Big) \nn \\ 
&\times& (\frac{1}{2}(Q^2+m_i^2-s) -\mD^2)-
       \frac{1}{s}(Q^2-\mD^2)(m_i^2-\mD^2)\ln F(Q^2)\Big]\Big\} \nn 
\ee
and the function $F(Q^2)$ has been defined above. 

The contribution from the double sneutrino graph is
\be
\sigma_i^{(\mbox{\footnotesize III,III})} & = &
         \frac{1}{2s}\frac{1}{4}\frac{1}{(2\pi)^5}
         \frac{1}{8} |V_{i1}|^2 \frac{g^6}{c_W^2}
         \frac{\pi^2}{3}s^2 
         \int d Q^2 \; 
         P(m_{Z^0},s) P(m_{\tilde{\nu}_l^{N'}},Q^2)
\nn \\ 
&\times& \lambda^{1/2}\Big( 1,\frac{Q^2}{s},\frac{\mNl^2}{s}\Big)    
         \lambda^{1/2}\big( 1,\frac{m_\ell^2}{Q^2},
                              \frac{m_{\chi_i}^2}{Q^2} \Big)
         (Q^2-m_{\chi_i}^2-m_\ell^2)
\nn \\ 
&\times& \Big[ 1-2\frac{\mNl^2}{s}-2\frac{Q^2}{s} -\frac{1}{2}
         \lambda\Big( 1,\frac{Q^2}{s},\frac{\mNl^2}{s} \Big) + 
         \frac{(m_{\tilde{\nu}^N_\ell}^2-Q^2)^2}{s^2} \Big] \ . \nn  
\ee
The integration limits are $Q^2_{min}=(m_\ell+m_{\chi_i})^2$
and $Q^2_{max}=(\sqrt{s}-m_{\tilde{\nu}_\ell^N})^2$. Note that $N \neq N'$
due to the off-diagonality of the $Z^0 \tilde{\nu}_\ell^1
\tilde{\nu}_\ell^2$ vertex. 


\section{Contributions to 
$\bm{\lowercase{e}^+ \gamma \ra \tilde{\nu}_\ell^N 
\tilde{\ell}^+ \ovl{\nu}_{\lowercase{e}}, \,  \ell=\mu,\tau}$}

The indivual amplitudes of the graphs shown in Fig.~\ref{snunuslep} are 
given by
\be
T^{(I)}_{\mu} &=& -D(s_2)D(t_1)j^{\alpha}
                  (p_-)_{\beta} 
                  (g^{\beta \nu}-\frac{Q^{\beta}Q^{\nu}}{m_W^2})
                  S_{\mu \nu \alpha} \ , \nn \\
T^{(II)}_{\mu} &=& D(s_2) (p_-)_{\alpha}
                  (g^{\alpha \nu} - \frac{Q^{\alpha} Q^{\nu}}{m_W^2})
                  \ovl{v}(p_e) \gamma_{\mu}
                  \frac{-(p_g \hskip-3.3mm /\; + p_e \hskip-3.2mm /\;)}{s}
                  \gamma_{\nu} P_L v(p_{\nu}) \ , \nn \\
T^{(III)}_{\mu} &=& -D(t_1) j_{\mu} \ , \nn \\
T^{(IV)}_{\mu} &=& D(t_1)\frac{1}{m_{\tilde{l}}^2-K^2}
                  j^{\alpha}[K_{\mu}K_{\alpha}+
                  (p_{\tilde{\nu}_l^N})_{\alpha}(K-p_{\tilde{l}})_{\mu}
                  -(p_{\tilde{l}})_{\mu} K_{\alpha}] \ . \nn 
\ee
The sum of the first and the third contribution can be simplified to give
\be
(T^{(I)}+T^{(III)})_{\mu}& = & 
                         D(s_2)D(t_1)j^{\alpha}p_-^{\beta} 2
                        (g_{\alpha \beta}P_{\mu}+
                         g_{\mu \alpha}(p_g)_{\beta}-
                         g_{\mu \beta}(p_g)_{\alpha}) \nn \\ 
                       & + &
                         D(s_2)\frac{-1}{s}p_-^{\alpha}
                         \ovl{v}(p_e)\gamma_{\mu}
                         (p_g \hskip-3.3mm /\; +p_e \hskip-3.2mm /\;)
                         \gamma_{\alpha} P_L v(p_{\nu}) \nn \ .
\ee
Here we have defined 
\be p_-=p_{\tilde{l}}-p_{\tilde{\nu}_l^N},\; 
    K=p_g-p_{\tilde{l}},\;
    Q= p_{\tilde{l}}+p_{\tilde{\nu}_l^N},\;
    P=p_e-p_{\nu},\;
    t_1=P^2,\;
    s_2=Q^2,\; \nn 
\ee

\vspace{-0.5cm}
\be
    j^{\alpha}=\ovl{v}(p_e)\gamma^{\alpha}P_L v(p_{\ovl{\nu}}), \;
    S_{\mu \nu \alpha}=-g_{\alpha \nu}(P+Q)_{\mu}
                      +g_{\mu \alpha}(P-p_g)_{\nu}
                      +g_{\mu \nu}(Q+p_g)_{\alpha},\nn
\ee

\vspace{-0.5cm}
\be
    D(s_2)=(m_W^2-s_2+i\Gamma_W m_W)^{-1},\;
    D(t_1)=(m_W^2-t_1)^{-1}\nn \ . \nn
\ee
Replacing in~\cite{singletop} the quark charges $e_t, e_b$ 
by 0, 1 respectively and the quark current $J_{quark}^{\mu}$ by 
$p_-^{\mu}=p_{\tilde{\ell}}-p_{\tilde{\nu}_\ell^N}$ 
(stemming from the derivative
$W\tilde{\ell}\tilde{\nu}_\ell$ coupling) leads to the equivalent expressions
of the amplitudes 
$T_{\lambda}^{(\mbox{\footnotesize III})},
 T_{\lambda}^{(\mbox{\footnotesize IV})}$ in 
~\cite{singletop}. A subtlety known from scalar electrodynamics
arises in the calculation of the time ordered product of the 
propagators of the exchanged scalar fields. Here the additional 
contribution from taking the partial derivatives out of the time 
ordered product is cancelled by the $W\gamma\tilde{\nu}_\ell\tilde{\ell}$
four point vertex. 


\section{Four-vector parametrization in terms of two invariants 
and two angles}

In order to carry out the numerical phase space integration in 
the process $e \gamma \ra \tilde{\ell}^+ \tilde{\nu}_\ell^N \ovl{\nu}_e$ 
following~\cite{Byk} the four-vectors in the process can be 
parametrized in terms of two invariants (see appendix C) and one solid 
angle as 
\be
p_{\gamma}  &=& \left(\begin{array}{c} 
     \dps{\frac{s-s_2}{2 \sqrt{s_2}}} 
     \Big[\sin^2\Theta+(\cos\Theta+
     {\dps \frac{s_2-t_1}{s-s_2}})^2\Big]^{1/2}\\
     \sin \Theta \dps{\frac{s-s_2}{2 \sqrt{s_2}}} \\
     0 \\
     {\dps \frac{s-s_2}{2 \sqrt{s_2}}}\Big[\cos \Theta 
                                      {\dps \frac{s_2-t_1}{s-s_2}}\Big]
     \end{array} \right) \ \ \ , \ \ \ 
p_e =  \left( \begin{array}{c} 
     {\dps \frac{s_2-t_1}{2 \sqrt{s}}} \\
     0 \\
     0 \\
     {\dps \frac{s_2-t_1}{2 \sqrt{s}}} \end{array}\right) 
     \ \ \ , \nn \\ \nn \\ \nn \\
p_{\nu} &=& \left( \begin{array}{c} 
     {\dps \frac{s-s_2}{2 \sqrt{s_2}}}\\
     \sin \Theta {\dps \frac{s-s_2}{2 \sqrt{s_2}}}\\
     0\\
     \cos \Theta {\dps \frac{s-s_2}{2 \sqrt{s_2}}} 
     \end{array}\right) \ \ \ , \ \ \ 
p_{\tilde{\ell}}  =  \left( \begin{array}{c}
     \Big[m_{\tilde{\ell}}^2+{\dps \frac{\lambda(s_2,m_{\tilde{\nu}_\ell^N}^2,m_{\tilde{\ell}}^2)}{4 s_2}}\Big]^{1/2}\\
     {\dps \frac{\lambda(s_2,m_{\tilde{\nu}_\ell^N}^2,m_{\tilde{\ell}}^2)^{1/2}}{2 \sqrt{s_2}}}
     \sin \theta \cos \phi \\
     {\dps \frac{\lambda(s_2,m_{\tilde{\nu}_\ell^N}^2,m_{\tilde{\ell}}^2)^{1/2}}{2 \sqrt{s_2}}}
     \sin \theta \sin \phi \\
     {\dps \frac{\lambda(s_2,m_{\tilde{\nu}_\ell^N}^2,m_{\tilde{\ell}}^2)^{1/2}}{2 \sqrt{s_2}}}
     \cos \theta 
     \end{array} \right) \nn \\ \nn \\
p_{\tilde{\nu}_\ell^N} &=& \left( \begin{array}{c}
     \Big[m_{\tilde{\nu}_\ell^N}^2 + {\dps \frac{\lambda(s_2,m_{\tilde{\nu}_\ell^N}^2,m_{\tilde{\ell}}^2)}{4 s_2}}\Big]^{1/2} \\
     - \vec{\bf{p}}_3
     \end{array}\right) \ \ , \nn \\ \nn \\ \nn \\
\cos \Theta &=& \frac{(s-s_2)(s_2-t_1)-2 s_2(s-s_2-t_1)}
                  {(s_2-t_1)(s-s_2)} \nn \ .
\ee

\newpage

\begin{table}[t]
\caption{Parameters used for numerical results assuming a gravity mediated 
GUT scheme. $V$ ($N$) is the chargino (neutralino) mixing matrix. 
}
\label{tablesets}
\begin{tabular}{cccc} 
\multicolumn{4}{c}
{\boldmath \mbox{\bf Set A:} 
$\mu=-100\, \hbox{ \bf GeV} \ \ , \ \ M_2=100\, \hbox{ \bf GeV}\ \ , \ \ 
\tan \beta = 2$} 
\\ \hline 
$m_{\chi^+_1}= 151.1 \hbox{ GeV}$ & $m_{\chi^+_2}=100.3 \hbox{ GeV} $ &
$V(1,1)=0.89$ & $V(1,2)=-0.44$ \cr 
$m_{\chi^0_1}=143.6 \hbox{ GeV}$ & $m_{\chi^0_2}=134.5 \hbox{ GeV}$ &
$m_{\chi^0_3}=86.2 \hbox{ GeV}$ & $m_{\chi^0_4}=54.8 \hbox{ GeV}$ \\ 
$N(1,1)=-0.20 $ & $N(2,1)=-0.21$ & $N(3,1)=-0.11$ & $N(4,1)=-0.95$ \\ 
$N(1,2)=0.78$ & $N(2,2)=0.30$ & $N(3,2)=-0.51$ & $N(4,2)=-0.18$\\ 
\hline \hline
\multicolumn{4}{c}
{\boldmath \mbox{\bf Set B:}
$\mu=-200\, \hbox{ \bf GeV}\ \ , \ \  M_2=200\, \hbox{ \bf GeV} \ \ ,\ \ 
\tan \beta = 35$} 
\\ \hline
$m_{\chi^+_1}= 263.5 \hbox{ GeV}$ & $m_{\chi^+_2}=153.2 \hbox{ GeV} $ &
$V(1,1)=0.80$ & $V(1,2)=-0.60$ \\ 
$m_{\chi^0_1}=262.2 \hbox{ GeV}$ & $m_{\chi^0_2}=211.7 \hbox{ GeV}$ &
$m_{\chi^0_3}=155.7 \hbox{ GeV}$ & $m_{\chi^0_4}=94.3 \hbox{ GeV}$ \\ 
$N(1,1)=0.15 $ & $N(2,1)=-0.10$ & $N(3,1)=-0.30$ & $N(4,1)=0.93$ \\ 
$N(1,2)=-0.70$ & $N(2,2)=0.14$ & $N(3,2)=-0.69$ & $N(4,2)=-0.10$
\end{tabular}
\end{table}

\ \ 

%
%
%
%

\newpage




\begin{figure} 
\caption{Like Sign Dilepton  via Drell-Yan sneutrino pair production 
at an electron or proton collider (for \(\ell \neq e\) ).}
\label{pairgraph}
\end{figure}


\begin{figure}
\caption{Contributions to 
$e^+ e^- \ra \tilde{\nu}_l^N \ell^- \tilde{\chi}^+_i$ 
where $N$=1,2 and $\ell$=$\mu,\tau$ (for $\ell$=$e$ more graphs 
contribute).}
\label{feyntc} 
\end{figure}



\begin{figure}
\caption{Contributions to 
$e^+\gamma \ra \tilde{\nu}_\ell^N \tilde{\ell}^+ \ovl{\nu}_e$. These are the
leptonic analogs to single top production in the SM. Graph II containing
a four point vertex has no SM-counterpart.}
\label{snunuslep}
\end{figure}



\begin{figure}
\caption{The L violating prefactor $\xi^{pair}$ in sneutrino
pair production defined in Eq.~(\ref{xifactor}). The solid lines
correspond to parameter set A and (from left to right) 
$\ovl m$=120 GeV, 170 GeV, 220 GeV, 300 GeV, 390 GeV; the dashed lines 
correspond to parameter set B and (from left to right) 
$\ovl m$=170 GeV, 220 GeV, 300 GeV, 390 GeV. The parameter
sets A and B are defined in Table~\ref{tablesets}.}
\label{pairxifig}
\end{figure}


\begin{figure}
\caption{LSD cross section at TEVATRON:
$\ovl m$=120 GeV, set A (solid line); 
$\ovl m$=170 GeV, set A (dashed line);
$\ovl m$=220 GeV, set A (dot-dashed line);
$\ovl m$=170 GeV, set B (double-dot dashed line);
$\ovl m$=220 GeV, set B (double-dash dotted line).
The parameter sets A and B are defined in Table~\ref{tablesets}.}
\label{lsdtevatron}
\end{figure}


\begin{figure}
\caption{LSD cross section at LHC for the parameter sets A (left) 
and B (right) defined in Table~\ref{tablesets} for the parameters
$\ovl m$=120 GeV, set A (solid line); 
$\ovl m$=170 GeV, set B (solid line);
$\ovl m$=200 GeV, sets A and B (dashed lines);
$\ovl m$=400 GeV, sets A and B (dot-dashed lines);
$\ovl m$=600 GeV, sets A and B (double dot-dashed lines).}
\label{lsdlhc}
\end{figure}


\begin{figure}
\caption{LSD cross section in sneutrino pair production
at a LC ($\ell$=$\mu, \tau$) for the parameters:
$\ovl m$=120 GeV, set A, $\sqrt{s}$=500 GeV (solid line);
$\ovl m$=170 GeV, set A, $\sqrt{s}$=500 GeV (dashed line);
$\ovl m$=220 GeV, set A, $\sqrt{s}$=500 GeV (dot-dashed line);
$\ovl m$=170 GeV, set B, $\sqrt{s}$=500 GeV (solid line);
$\ovl m$=220 GeV, set B, $\sqrt{s}$=500 GeV (dashed line);
$\ovl m$=120 GeV, set A, $\sqrt{s}$=800 GeV (solid line);
$\ovl m$=220 GeV, set A, $\sqrt{s}$=800 GeV (dashed line);
$\ovl m$=300 GeV, set A, $\sqrt{s}$=800 GeV (dot-dashed line);
$\ovl m$=390 GeV, set A, $\sqrt{s}$=800 GeV (double dot-dashed line);
$\ovl m$=170 GeV, set B, $\sqrt{s}$=800 GeV (solid line);
$\ovl m$=220 GeV, set B, $\sqrt{s}$=800 GeV (dashed line);
$\ovl m$=300 GeV, set B, $\sqrt{s}$=800 GeV (dot-dashed line);
$\ovl m$=390 GeV, set B, $\sqrt{s}$=800 GeV (double dot-dashed line).
The parameter sets A and B are defined in Table~\ref{tablesets}.}
\label{lsdnlc}
\end{figure}


\begin{figure}
\caption{
Cross section for chargino-associated sneutrino production depicted 
in Fig.~\ref{feyntc}-I and~\ref{feyntc}-II as a function of the
sneutrino mass. The curves correspond to the parameters
($\sqrt{s}$=500 GeV; parameter set A; lower solid line); 
($\sqrt{s}$=800 GeV; set A; upper solid line);
($\sqrt{s}$=500 GeV; set B; lower dashed line);
($\sqrt{s}$=800 GeV; set A; upper dashed line). 
The parameter sets A and B have been defined in Table~\ref{tablesets}.
The sneutrino has been taken to be heavier than the charginos,
otherwise the production of two real charginos is allowed and the
chargino width must be taken into account.}
\label{mass2char}
\end{figure}



\begin{figure}
\caption{The L-violating parameter in single sneutrino production
$\xi^{single}$ defined in Eq.~(\ref{xidef}). The varius curves refer 
to the  parameters used in Fig.~\ref{loglogfin} with the same notation.}
\label{xifigure}
\end{figure}



\begin{figure}
\caption{Cross section for LSD
stemming from the cha\-rgi\-no-as\-so\-cia\-ted sneutrino production depicted 
in Fig.~\ref{feyntc}-I and~\ref{feyntc}-II. 
The curves correspond to the parameters
$\sqrt{s}=$500 GeV, $\overline{m}$=275 GeV (parameter set A, 
                                             upper solid line); 
$\sqrt{s}=$500 GeV, $\overline{m}$=350 GeV (set A, lower solid line); 
$\sqrt{s}=$800 GeV, $\overline{m}$=450 GeV (set A, upper dashed line);
$\sqrt{s}=$800 GeV, $\overline{m}$=600 GeV (set A, lower dashed line);
$\sqrt{s}=$800 GeV, $\overline{m}$=450 GeV (set B, upper dotted line); 
$\sqrt{s}=$800 GeV, $\overline{m}$=600 GeV (set B, lower dotted line).
The parameter sets A and B have been defined in Table~\ref{tablesets}.}
\label{loglogfin}
\end{figure}




\begin{figure}
\caption{cross section for 
$e^+\gamma \ra \tilde{\nu}_\ell^N \tilde{\ell}^+ \ovl{\nu}_e$
in a mSUGRA scenario as a function of the common scalar mass 
$M_0$, applying the
backscattering process (B.S.) Eq.~(\ref{foldedcross}) and the equivalent 
photon approximation (E.P.A) Eq.~(\ref{photonapprox}). The curves 
correspond to the parameters
($\sqrt{s}$=800 GeV, parameter set A; E.P.S.; upper solid line); 
($\sqrt{s}$=500 GeV, set A; E.P.S.; lower solid line); 
($\sqrt{s}$=800 GeV, set B; E.P.S.; upper long-dashed line);
($\sqrt{s}$=500 GeV, set B; E.P.S.; lower long-dashed line);
($\sqrt{s}$=800 GeV, set A; B.S.; upper dashed line); 
($\sqrt{s}$=500 GeV, set A; B.S.; lower dashed line);
($\sqrt{s}$=800 GeV, set B; B.S.; upper dotted line); 
($\sqrt{s}$=500 GeV, set B; B.S.; lower dotted line).
The parameter sets A and B have been defined in Table~\ref{tablesets}
(the center of mass energy refers to the electron-positron beam).}
\label{bare}
\end{figure}


   
\begin{figure}
\caption{cross section for the wrong sign charged lepton signal
stemming from
$e^+\gamma \ra \tilde{\nu}_\ell^N \tilde{\ell}^+ \ovl{\nu}_e$
(in a mSUGRA scenario) as a function of the ratio 
$\Delta m/\overline{m}$ applying the
backscattering process (B.S.) Eq.~(\ref{foldedcross}) and the equivalent 
photon approximation (E.P.A) Eq.~(\ref{photonapprox}). The curves 
correspond to
($\sqrt{s}$=500 GeV, $M_0$=100 GeV; parameter set A; E.P.S.; 
                                             upper solid line); 
($\sqrt{s}$=500 GeV, $M_0$=100 GeV; set A; B.S.; lower solid line); 
($\sqrt{s}$=800 GeV, $M_0$=300 GeV; set A; E.P.S; upper dashed line);
($\sqrt{s}$=800 GeV, $M_0$=300 GeV; set A; B.S.; lower dashed line);
($\sqrt{s}$=800 GeV, $M_0$=300 GeV; set B; E.P.S; upper dotted line); 
($\sqrt{s}$=800 GeV, $M_0$=300 GeV; set B; B.S.; lower dotted line).
The parameter sets A and B have been defined in Table~\ref{tablesets}.
For $M_0$=100 GeV the data is cut off where the lighter sneutrino
becomes lighter than the LSP, the bump corresponds to the value
of $\Delta m$ where the heavier sneutrino is not produced any more.
}
\label{egamma}
\end{figure}



\begin{figure}
\caption{cross section for the wrong sign charged lepton signal
stemming from
$e^+\gamma \ra \tilde{\nu}_\ell^N \tilde{\ell}^+ \ovl{\nu}_e$
as a function of the ratio $\Delta m/\overline{m}$ applying the
backscattering process (B.S.) Eq.~(\ref{foldedcross}).  
No mSUGRA conditions were assumed. The curves correspond to
($\sqrt{s}$=500 GeV, $\ovl{m}$=200 GeV; $m_{\tilde{\ell}}$=100 GeV;
parameter set A; B.S.; solid line);
($\sqrt{s}$=800 GeV, $\ovl{m}$=300 GeV; $m_{\tilde{\ell}}$=200 GeV;
parameter set A; B.S.; dashed line);
The parameter set A has been defined in Table~\ref{tablesets}.
}
\label{noSUGRA}
\end{figure}


\begin{figure}
\caption{Cross section of the LSD signal in single sneutrino
production at an electron collider Eq.~(\ref{crosslviol}) for the case
of a very small common sneutrino width 
$\Gamma:=$$\Gamma_1$=$\Gamma_2$=
$\Gamma_1(\chi^{\pm} \ell^{\mp})$=
$\Gamma_2(\chi^{\pm} \ell^{\mp})$ and for the
and for the
parameters (set A; $\sqrt{s}$=500 GeV; $\overline{m}$=275 GeV) in 
dependence of $\Delta m$ (in GeV). The solid line corresponds to
$\Gamma$=10 MeV, the dashed line corresponds to $\Gamma$=1 MeV and
the dot-dashed line to $\Gamma$=100 keV. For a neutrino mass of order
${\cal O}$(1 eV) ($\Delta m \, \approx \, 200$ keV $=2\times 10^{-4}$ GeV, 
see Eq.~\ref{neutrlimits})
the LSD
is of order ${\cal O}$(0.01 fb) (dashed line).}
\label{varywidth}
\end{figure}


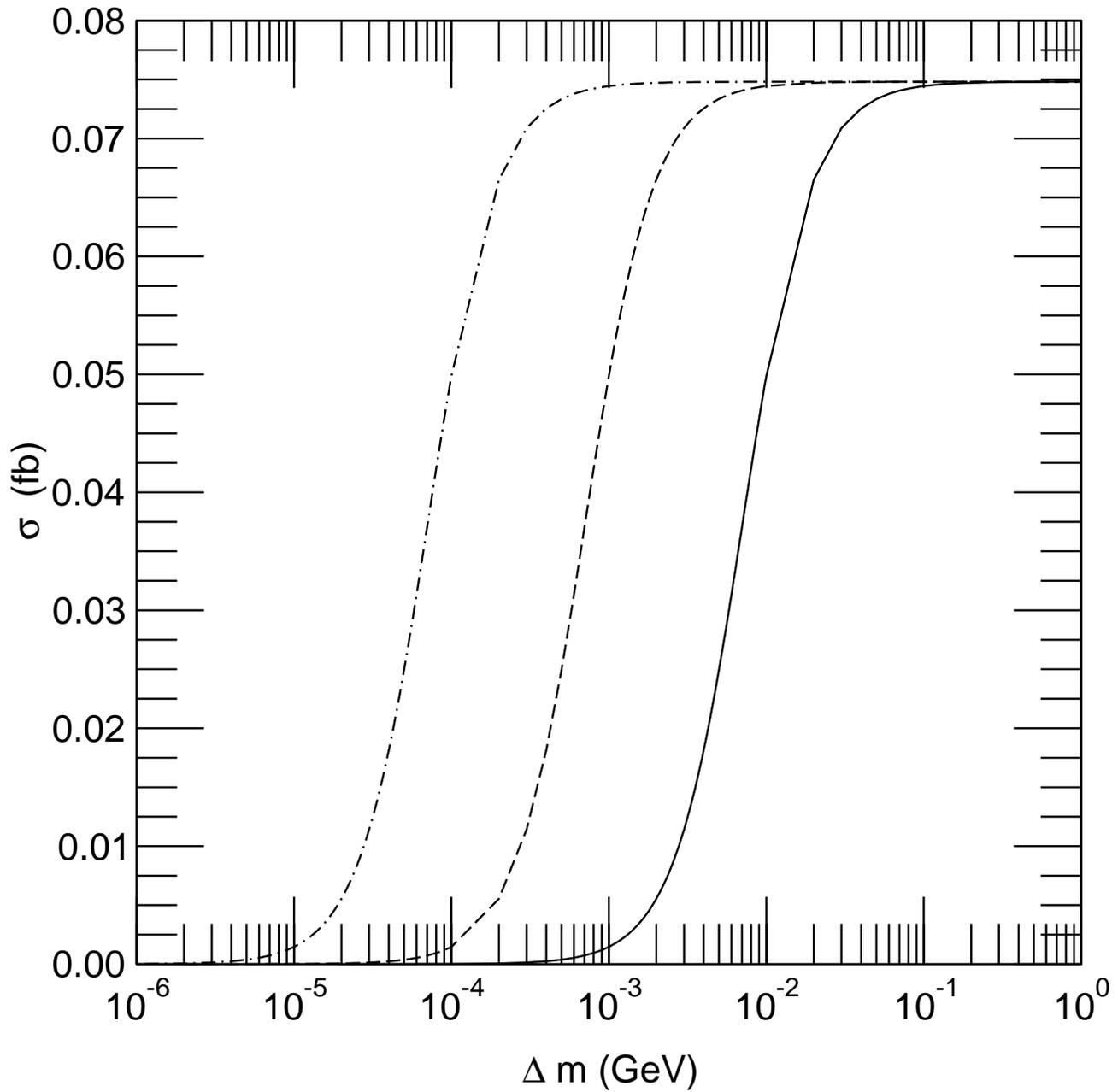
\end{document}